\documentclass[journal]{IEEEtranTCOM}
\normalsize
\usepackage[pdftex]{graphicx}
  \usepackage{epstopdf}
  \DeclareGraphicsExtensions{.pdf,.jpeg,.png,.eps}
\usepackage{color,balance}
\usepackage[cmex10]{amsmath}
\usepackage{amssymb}
\usepackage{algorithmic}
\usepackage{array}
\usepackage{mdwmath}
\usepackage{amsthm}
\usepackage{multirow}
\usepackage{booktabs}
\usepackage{comment}
\usepackage{graphicx}
\usepackage{textgreek}
\usepackage{float}
\usepackage{cite}
\usepackage{fixltx2e}
\usepackage{amsmath}
\usepackage{bm}

\usepackage{mdwtab}
\usepackage{tabularx}
\usepackage{xcolor}
\usepackage{adjustbox}
\usepackage{balance}
\usepackage{cite}
\makeatother

\allowdisplaybreaks
\newtheoremstyle{mystyle}
  {3pt}
  {3pt}
  {\itshape} 
  {\parindent}
  {\bfseries}
  {\upshape{:}}
  {.5em}
  {}

\usepackage{color,soul}
\usepackage{setspace}
\theoremstyle{mystyle}

\theoremstyle{mystyle}  

\theoremstyle{mystyle}

\usepackage[british,UKenglish]{babel}
\usepackage{subcaption}
\usepackage[font={small}]{caption}
\graphicspath{{./figures/}}

\usepackage{tabularx}
\usepackage{booktabs}
\usepackage{array}

\begin{document}

\title{Intelligent Blockage Prediction and Proactive Handover for Seamless Connectivity in Vision-Aided 5G/6G UDNs}

\author{Mohammad Al-Quraan,~\IEEEmembership{Student Member,~IEEE,} Ahsan Khan, Lina Mohjazi,~\IEEEmembership{Senior Member,~IEEE,} Anthony Centeno, Ahmed Zoha,~\IEEEmembership{Member,~IEEE,} and Muhammad Ali Imran,~\IEEEmembership{Senior Member,~IEEE}

\thanks{M. Al-Quraan, A. Khan, L. Mohjazi, A. Centeno, A. Zoha, and M. A. Imran are with the James Watt School of Engineering, University of Glasgow, Glasgow, G12 8QQ, UK. (e-mail: \{m.alquraan.1, a.khan.9\}@research.glasgow.ac.uk, \{Lina.Mohjazi, Anthony.Centeno, Ahmed.Zoha, Muhammad.Imran\}@glasgow.ac.uk).}
}

\maketitle
\markboth{}{}
\begin{abstract}
The upsurge in wireless devices and real-time service demands force the move to a higher frequency spectrum. Millimetre-wave (mmWave) and terahertz (THz) bands combined with the beamforming technology offer significant performance enhancements for ultra-dense networks (UDNs). Unfortunately, shrinking cell coverage and severe penetration loss experienced at higher spectrum render mobility management a critical issue in UDNs, especially optimizing beam blockages and frequent handover (HO). Mobility management challenges have become prevalent in city centres and urban areas. To address this, we propose a novel mechanism driven by exploiting wireless signals and on-road surveillance systems to intelligently predict possible blockages in advance and perform timely HO. This paper employs computer vision (CV) to determine obstacles and users' location and speed. In addition, \textit{this study introduces a new HO event, called block event (\textbf{BLK}), defined by the presence of a blocking object and a user moving towards the blocked area}. Moreover, the multivariate regression technique predicts the remaining time until the user reaches the blocked area, hence determining best HO decision. Compared to typical wireless networks without blockage prediction, simulation results show that our BLK detection and PHO algorithm achieves 40\% improvement in maintaining user connectivity and the required quality of experience (QoE).

\end{abstract}
\begin{IEEEkeywords}
Computer vision, object detection, machine learning, blockage prediction, proactive handover, mobility management, mmWave communications, ultra-dense networks.
\end{IEEEkeywords}

\section{INTRODUCTION}
Millimetre-wave (mmWave) and terahertz (THz) technologies are vital in supporting beyond fifth-generation (B5G) and sixth-generation (6G) networks. The dependence on new high-frequency bands is expected to achieve the global connectivity vision by providing significant enhancements in terms of multi-Gbit/s throughput, supporting a massive number of devices, and delivering ultra-low latency and reliable connections \cite{[J.18]}. Moreover, the transition towards higher bands changes the paradigm of future wireless networks to small coverage cells, and thus, forming the concept of ultra-dense networks (UDNs) \cite{[J.19]}. UDNs are needed to meet the stringent broadband access demands and realise various revolutionary applications, such as intelligent healthcare, holographic telepresence, and autonomous driving.

UDNs leverage mmWave and THz multiarray antennas that offer beamforming capabilities, which focus the power of radio signal towards the receiving device through line-of-sight (LoS) communication. Beamforming is expected to be extensively used in next-generation networks owing to the provision of remarkable features, like high spatial reuse, increased throughput, boosting capacity, and interference elimination \cite{[J.20]}. Despite the outstanding merits offered by the UDNs, relying on high-frequency beam-based communications is more sensitive to the adverse effects of blockage and penetration losses than those operating at lower frequency bands. For instance, a link budget undergoes a 20 dB or more power loss when the connection is blocked by obstacles, such as human bodies or vehicles \cite{[J.21],[J.22]}. Such a sudden drop in the received power affects the signal quality and degrades the data rate of the communication link, making the network unreliable for time-sensitive applications.

Network densification aims to serve the highly populated urban areas and meet user demands and traffic capacity. However, cell coverage shrinkage and the presence of dynamic users/obstacles means mobility management in UDN is far more complex than in legacy networks, resulting in the recurrence of challenging issues, namely beam blockage and frequent handover (HO) \cite{[J.25]}. HO is a fundamental mechanism in any wireless network that transfers the ongoing call or data session from one base station (BS) to another. The 3rd Generation Partnership Project (3GPP) organisation introduced several predefined measurement events; if one occurred, HO must be conducted \cite{[J.26]}. Typically, a user equipment (UE)-assisted network for controlling HO receives a measurement report from the user with information about the received signal strength (RSS)/quality of a specific downlink reference signal from the serving BS (S-BS) and other neighbouring BSs. If the condition of a specific event is met, the network will trigger the HO process and negotiations eventuate between the S-BS and the target BS (T-BS) to complete handing the user to the new BS, thus guaranteeing user connectivity.

Unlike in 3G/4G networks, mobility management in B5G networks is accompanied with negative impacts at both user and network levels. Frequent HOs are leading to more data transmission delays and throughput loss in the UE as well as causing increased power consumption and poor network quality of service (QoS), especially if there is rejection from the T-BSs due to full resource occupation. This problem is exacerbated in smart cities due to the highly dynamic environment and the existence of blocking objects that can shade the serving beam. Selection/reselection of the best beam is a time-consuming task that will result in additional network operating costs. Accordingly, beam blockage and frequent HO are attracting the attention of research bodies in academia and industry to find new solutions that can improve the reliability of UDNs.

\subsection{Related Work}
Despite numerous benefits gained when shifting operational frequencies from the lower bands (sub-6GHz) to the higher bands (mmWave and THz), reliance on mmWave and THz technologies introduces critical challenges, such as link blockage and frequent HOs. To this end, many attempts have been made to provide solutions to address the connectivity issue in mmWave networks. In \cite{[J.5]}, the authors count on the geometry of mmWave channels to predict when and for how much time an LoS link will be blocked by observing the behaviour of the neighbouring non-LoS (NLoS) signals. The main idea is that a connected user will be served by LoS and NLoS links, and detecting a blockage in one of the NLoS connections can be exploited to anticipate when the LoS link will be blocked, thus performing HO proactively. However, such techniques are not efficient in practical scenarios due to assuming slowly moving obstacles, highly scattering environment, and NLoS will be blocked before the LoS links. The work in \cite{[J.6]} explores the use of channel state information of the sub-6GHz channels and the effect of frequency-dependent diffraction to form an early warning of possible mmWave signal blockages in hybrid communication systems. Motivated by the fact that the diffraction angle decreases as the frequency increases, the diffracted sub-6GHz signals reach a certain signal strength threshold earlier than the mmWave signals. This work relies on simulations to validate the proposed method, while the authors did not measure the sub-6GHz and mmWave diffraction in reality to identify a significant difference. Liu and Xiao \cite{[J.8]} followed a different mechanism to predict beam blockage in heterogeneous mmWave networks by observing the previous beamforming vectors and their fingerprints. The cloud radio access network (C-RAN) is used to maintain a fingerprinting database that can be used to predict possible future blockages and apply the needed countermeasures in advance. The drawback of this mechanism is that it takes a long time to build the fingerprinting database table that also needs to be updated frequently and therefore, it is not suitable for dynamic environments.

Subsequently, learning-based approaches have exploited various ML techniques to optimise the operations of UDNs. In the interest of maintaining seamless connectivity and fulfilling the required QoS for mobile users, the work in \cite{[J.1]} presents a proactive mobility management scheme. This work comprises a resource reservation and prediction scheme, which uses neural network (NN) to predict the average channel quality and the connected BSs for each real-time user, hence reserving the required resources and guaranteeing the users’ QoS requirements. Besides, a proactive HO (PHO) scheme, which aims to avoid frequent HOs and reduce HO latency by relying on the dual connectivity (DC) mechanism. The DC allows each user to connect with more than one BS simultaneously to achieve zero HO interruption time and maintain seamless connectivity. Likewise, the study in \cite{[J.2]} exploits the DC and deep learning (DL) algorithm to avoid service interruption during HO decisions. A long short-term memory (LSTM) model is trained to predict the user’s future movement trends depending on historical trajectory information to perform efficient HOs. Despite the potential of eliminating the HO intermittency, the DC technique will add more operational complexity to the network as well as to user devices. In addition to incurring more costs, wasting network resources, and increasing energy consumption. Moreover, multiconnectivity will not avoid LoS links blockages and service disconnection. Based on dual band network operation, the authors in \cite{[J.4]} exploit the knowledge of the sub-6GHz uplink channel to enhance the reliability of the mmWave downlink channels motivated by the spatial correlation between the two frequency bands. A DL model is trained using a tuple of sub-6GHz channel information and blocking status to determine whether the LoS link is blocked or not. However, this work would not suit realistic scenarios since it only classifies the channel status as blocked/unblocked and cannot avoid blockages in advance.

The emerging research direction of exploiting computer vision (CV) for developing wireless communications and tackling complicated problems in UDNs has gained much interest recently. The consistency between the LoS communication and the direct camera view is envisioned to play a major role in future wireless networks. For instance, the study in \cite{[J.9]} leverages camera imagery and DL to tackle the beam blockage problem in mmWave systems. The proposed technique predicts the time series of the mmWave received power to several hundred milliseconds in advance based on the depth images of the served area, allowing for sufficient time to perform HO. However, predicting the received power in advance does not necessarily suit HO decision problems. Furthermore, it requires large quantities of training datasets and computational resources to prepare a model for accurate prediction. The authors in \cite{[J.10]} utilise the visual sensory information collected from the served area to train a DL model and predict beam blockages. RGB images captured by a camera installed on each BS are labelled with the beam blockage status and used to train the ResNet18 model to classify images based on blockage status. However, the proposed technique does not predict in advance, so service disconnection can not be avoided.

\subsection{Motivation and Contribution}
As discussed in the previous section, the studies mentioned above have different assumptions that limit their applicability in real practical scenarios. The vision for B5G and 6G networks is to meet the strict requirements of maintaining high levels of QoS/quality of experience (QoE) and ensuring user connectivity to realise real-time services and applications. Therefore, the newly emerging research direction of exploiting CV to enhance the performance of mmWave communication systems is envisioned to assist the operation of such systems, satisfy the stringent demands, and encourage their widespread. However, merging CV in the operation of UDNs is still in its infancy and needs a lot of research devotion to getting the most out of it.

In this paper, we propose a novel CV-assisted PHO mechanism that combines two modes of information, i.e., wireless and imagery information, to predict possible beam blockages in advance and then instructs the network to perform HO in a time that maximises the overall quality of experience (QoE), hence optimising the network performance. An object detection and localisation (ODL) algorithm is adopted to analyse the RGB images from vision sensors, detect obstacles and users, and determine their location. Additionally, a simple NN model is trained using the multivariate regression method to predict the remaining time until the user is being blocked by the obstacle. This work introduces a new HO event, called blocking event (BLK), defined by the existence of a blocking object and a user moving towards the blocked area. Once a BLK event is detected, the proposed algorithm determines the best time to trigger HO and switch the user to another BS, therefore improving the network's reliability. The main contributions of this paper are as follows:
\begin{itemize}
    \item First, we present a novel solution to the problem of beam blockage and frequent HOs in UDNs by utilising CV and NN algorithms. The CV is used to increase the network's awareness of the surrounding environment, and the NN model predicts when a sudden drop in the RSS will happen due to the existence of stationary obstacles, which is a very common challenge in UDNs.
    
    \item Furthermore, we introduce a new HO event called BLK, which can be considered in B5G and 6G networks besides the standardised events defined by the 3GPP \cite{[J.26]}. The BLK event is defined by detecting the presence of an obstacle and a user moving toward the blocked area.
    
    \item To determine the best point of triggering and completing HO once the BLK event is detected, this study provides an analysis on determining the optimal HO trigger point to keep the user’s QoE at a high level.
    
    \item Finally, we validate the accuracy of the proposed framework using state-of-the-art simulation tools. The results demonstrate the significance of this solution in maintaining seamless connectivity.
    
\end{itemize}

\subsection{Organisation}
The rest of this paper is organised as follows. Section \ref{Framework} presents the scenario under study and the schematic diagram in addition to describing the proposed framework in detail. In Section \ref{PEval}, performance evaluation and simulation results are discussed. Finally, Section \ref{Conclusion} gives concluding remarks.

\section{PROPOSED CV-Assisted PHO FRAMEWORK} \label{Framework}

The key idea of this work is to anticipate future beam blockage using CV and NN to perform timely PHO. Beam blockage prediction is a very challenging task because it depends on finding the location of a moving user and its possible sources of blockage in a realistic wireless scenario. In CV, ODL is used to identify an object's class and location coordinates; however, object detection alone is not enough to determine future blockages that necessitate: first, an efficient system that can detect the moving users (wireless users) and the potential source of blockage. The second is extracting augmented information, including speed, time, and distance from the blocked area. Guided by the above notions, the beam blockage prediction is divided into two sub-tasks, (i) ODL, which is used to determine the types of objects and their location to calculate their speed, (ii) Using the multivariate regression model to predict the remaining time until the users reaches the blocked area based on the information extracted from the RGB images. Subsequently, we will discuss in detail the various components of the proposed framework.

\subsection{The Scenario under Study}
\begin{figure}
\centering
\includegraphics[scale=0.32]{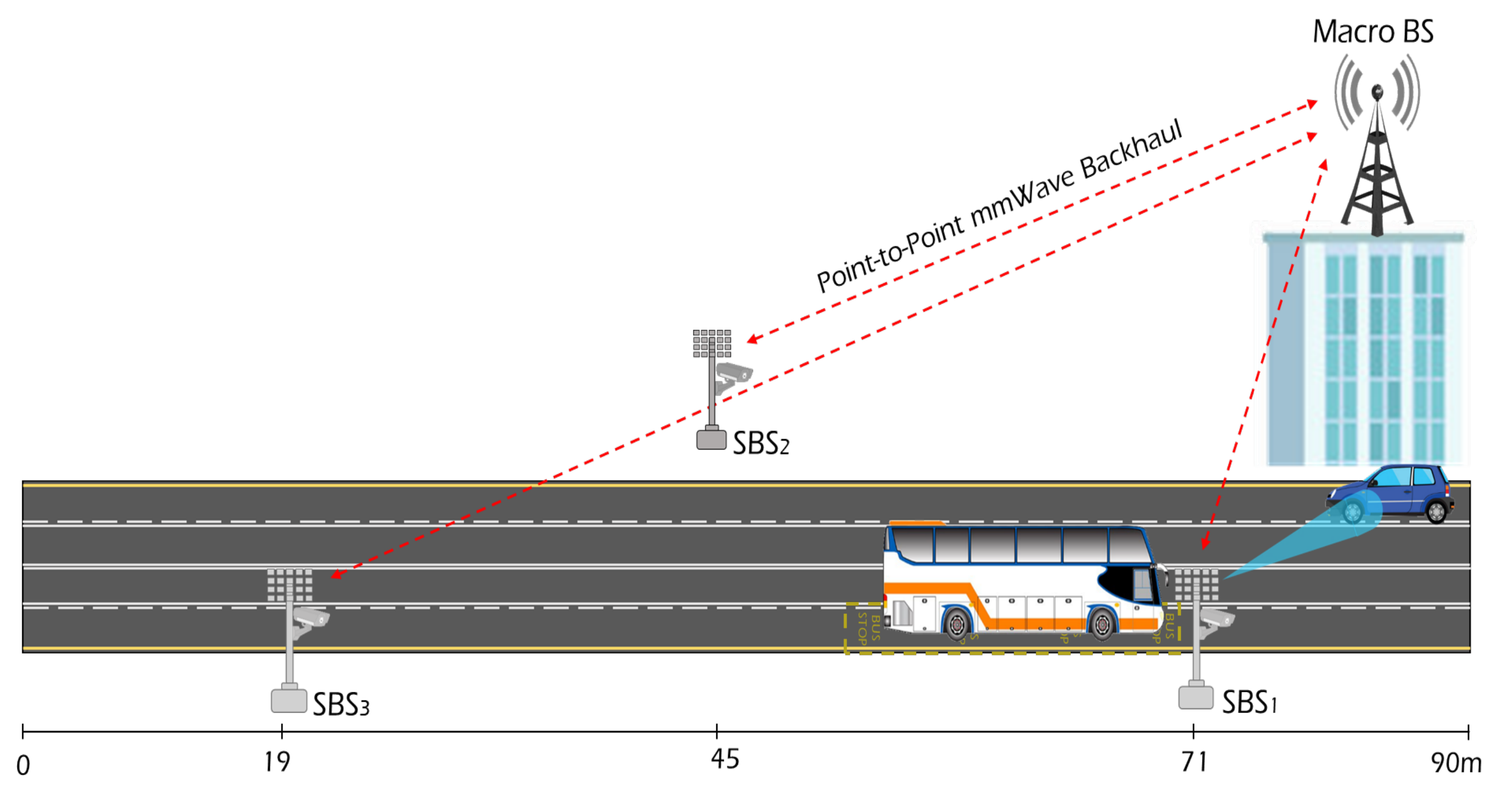}
\caption{The proposed system model: UDN including one macro BS and three SBSs each equipped with an RGB camera.}
\label{SM}
\end{figure}

We consider a wireless communication system consisting of one macro BS and three SBSs covering a 90$\times$15m street, as illustrated in Fig. \ref{SM}. The system adopts orthogonal frequency division multiplexing (OFDM) and operates at 60 GHz. Each SBS is equipped with an array of antennas that enables beamforming technology to create LoS beams that can achieve high RSS at a single-antenna user. Moreover, each SBS has a vision sensor (RGB camera) that captures visual information from the covered area to assist the operation of the UDN. It is worth mentioning that cameras are widely available for surveillance, especially in crowded places, which are the main target of deploying UDNs. The same surveillance cameras can be exploited for our system, hence reducing installation and operating costs.

The vision information is transmitted to a central server located at the macro BS through 10Gbps point-to-point mmWave backhaul links \cite{[J.24]}. The role of the central server is to collect, process, and use the visual information to train an ML model that can predict possible beam blockages in advance. This paper assumes a simple scenario including a moving user (vehicle) and a stationary blocking object (bus) that blocks the LoS communication between the SBS and the user.

\subsection{Schematic Diagram of the Proposed Framework}
\begin{figure}
\centering
\includegraphics[scale=0.344]{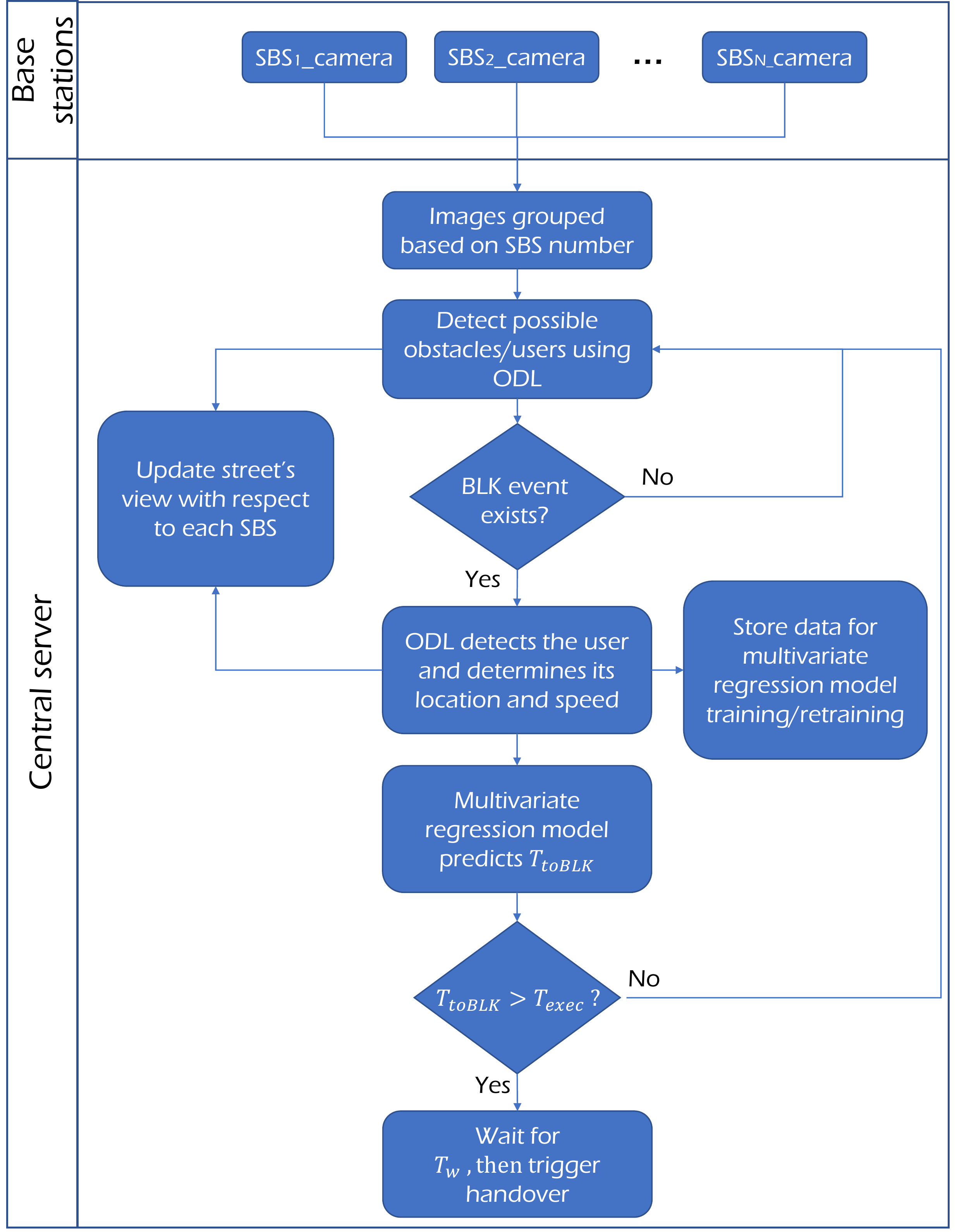}
\caption{Schematic diagram of our proposed framework.}
\label{SD}
\end{figure}

This study aims to achieve a blockage prediction mechanism, such that the network can proactively HO the user well before it reaches the blocked area. \textit{Once a BLK event is spotted out in the camera's field of view, the algorithm's main task is to predict the time needed by the user to reach the shadowed area, denoted as $\bm{T}_{\bm{toBLK}}$}. This time allows us to determine the best instant to perform HO before the user reaches that area and undergoes service interruption. Fig. \ref{SD} demonstrates the schematic diagram of the proposed technique. Multivariate regression is used to predict the $T_{toBLK}$ by modelling and training a two-hidden layer NN. Initially, the server will build a complete view of the covered area (the street) with the exact coordinates and locations of each SBS. The RGB cameras continuously\footnote{Note that some cameras have a motion detection feature that can be activated to reduce the amount of vision information sent to the server \cite{[J.27]}.} capture images from the covered area, then every SBS adds its identification number and timestamp to each image before sending it to the central server through the mmWave backhaul link. Once received, the server performs the following tasks:
\begin{itemize}
    \item First, it uses the ODL algorithm to detect blockages/users and updates its view. If a BLK event is detected, the server will move to the next step; otherwise, it will return to the detection phase.
    \item Then, the server identifies the user's exact location and updates its view, the user’s location information and the timestamp difference between two consecutive images are used to determine the user’s speed.
    \item After that, the location and speed information are stored for model training/retraining and used to predict the $T_{toBLK}$.
    \item Finally, if the $T_{toBLK}$ is greater than the execution time of the proposed algorithm ($T_{exec}$), the server will wait for a specific time and then send a HO trigger event to the network in order to HO the user to another SBS. Otherwise, it will return to the detection phase.
\end{itemize} 

\textit{The {$\bm{T}_{\bm{exec}}$} is defined as the time required by the proposed algorithm to be completed, starting from when the RGB images are captured until the HO process is completed}. It is divided into five sub times: (i) The time required to send two consecutive RGB images to the central server ($T_{RGB}$), (ii) the time needed to perform ODL on the two images ($T_{ODL}$), (iii) regression model inference time ($T_{inf}$), (iv) the waiting time after completing regression inference and before triggering HO ($T_{w}$), and (v) HO implementation time ($T_{HO}$).
\begin{equation}
    T_{exec} = T_{RGB} + T_{ODL} + T_{inf} + T_{w} + T_{HO}.
    \label{TExec}
\end{equation}
It is noteworthy that the values of all parameters specified in (\ref{TExec}) are fixed except for the value of $T_{w}$, which may vary based on the user's location. Determining the values of $T_{RGB}$, $T_{ODL}$, $T_{inf}$, and $T_{HO}$ depends on the capacity of the mmWave backhaul links, the type of model used, and the specifications of the central server. While choosing the value of $T_{w}$ is related to defining the optimal HO trigger region where the server will find out the optimal time to perform HO. 

\subsection {Object Detection and Localisation (ODL)}\label{ODETL}
To detect the objects' presence, position, and speed, the central server needs to receive at least two consecutive images (frames) from each SBS. Assuming that each camera records vision information at 26 fps \cite{[J.23]}, the time required to transmit these images from the SBS to the server ($T_{RGB}$) over 10 Gbps mmWave backhaul links will be very small (i.e., in microseconds) if each image is one megabyte in size.

\begin{figure}
\centering
\includegraphics[scale=0.37]{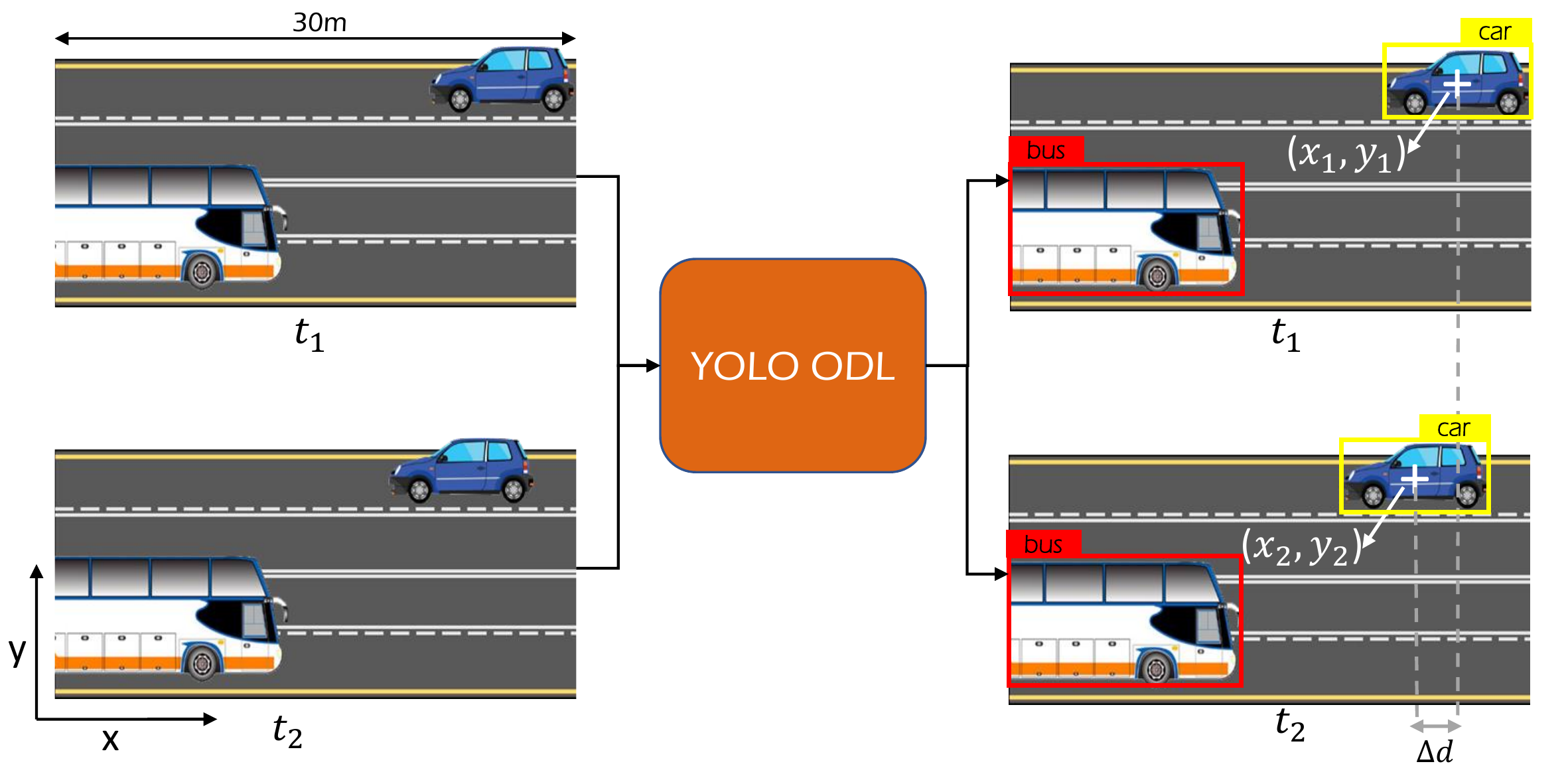}
\caption{Using ODL to detect objects and determine their locations. This information is used to determine the speed of the moving object.}
\label{Yolo}
\end{figure}

Once the server receives the visual information, the first step is to process that information to obtain the location of the objects. In this work, we adopt a state-of-the-art detection model, you only look once (YOLO) version 3, which can provide fast and accurate real-time object detection \cite{[J.14]}. Instead of developing and training an object detection model from scratch, YOLO models can be used directly without the need for any modification. Furthermore, the main objective of ODL is to identify if a BLK event exists and determine the objects’ locations in the pixel scale, then this information will be converted to the meter scale to find the speed of the objects. The server will feed the RGB images to the object detection algorithm, which in its turn detects the objects within the image by drawing bounding boxes around them and adding tags showing their categories, as illustrated in Fig. \ref{Yolo}. Moreover, this algorithm will provide the location information of the objects by determining the coordinates of the upper left and lower right corners of the bounding boxes. We use this information to determine the centre of the moving user, as shown in Fig. \ref{Yolo}.

The location information obtained from the YOLOv3 model is in pixel scale, and to determine the user's speed, we need to convert this information into a metric scale. The cameras' field of view is set to 100 degrees to ensure that the entire street is covered with minimal overlapping. Since the cameras capture images in two dimensions only, we assume that the image is flat and its width is equal to 30m distance\footnote{Determining image width and height in meter scale can be easily performed in real-world systems by capturing an image from the camera and measuring the actual image dimensions based on image corners. }. Therefore, the following formula can be used to determine the user's displacement in metres:
\begin{equation}
Travelled\_distance =  \frac{W_m}{W_p} \times \Delta d,
\end{equation}
where $W_{m}$ is the image width in meters, $W_{p}$ is the image width in pixels, and $\Delta d = |x_1 - x_2|$ is the x-axis user displacement in the pixel scale, assuming the user is moving in a straight trajectory, as demonstrated in Fig. \ref{Yolo}. For example, if a camera produces images with a resolution of 640$\times$480, then every 21 pixels are approximately equal to a distance of one meter. After evaluating the travelled distance, it is necessary to determine the associated travel time, which can be easily measured by exploiting the timestamp information of each image. Since the camera's frame rate is 26 fps, the time difference between two consecutive images will be about 38.5 ms. Now, the following speed formula can be used to determine the speed of the moving user:
\begin{equation}
    Speed = \frac{Distance}{Time}.
    \label{speed}
\end{equation}

Since the proposed solution depends on the $T_{exec}$ to determine if we have sufficient time to perform PHO, as demonstrated in Fig. \ref{SD}, we need to find out the detection time of the YOLOv3 model ($T_{ODL}$). Benefiting from the high performance of multi-access edge computing (MEC) servers, the $T_{ODL}$ can be reduced significantly to tens of ms when using the MEC server as a central server. Based on the analysis presented in \cite{[J.14]}, we assume that the $T_{ODL}$ requires 102 ms for detecting objects in two images.

\subsection{Multivariate Regression: Learning and Prediction}
\begin{figure}
\centering
\includegraphics[scale=0.42]{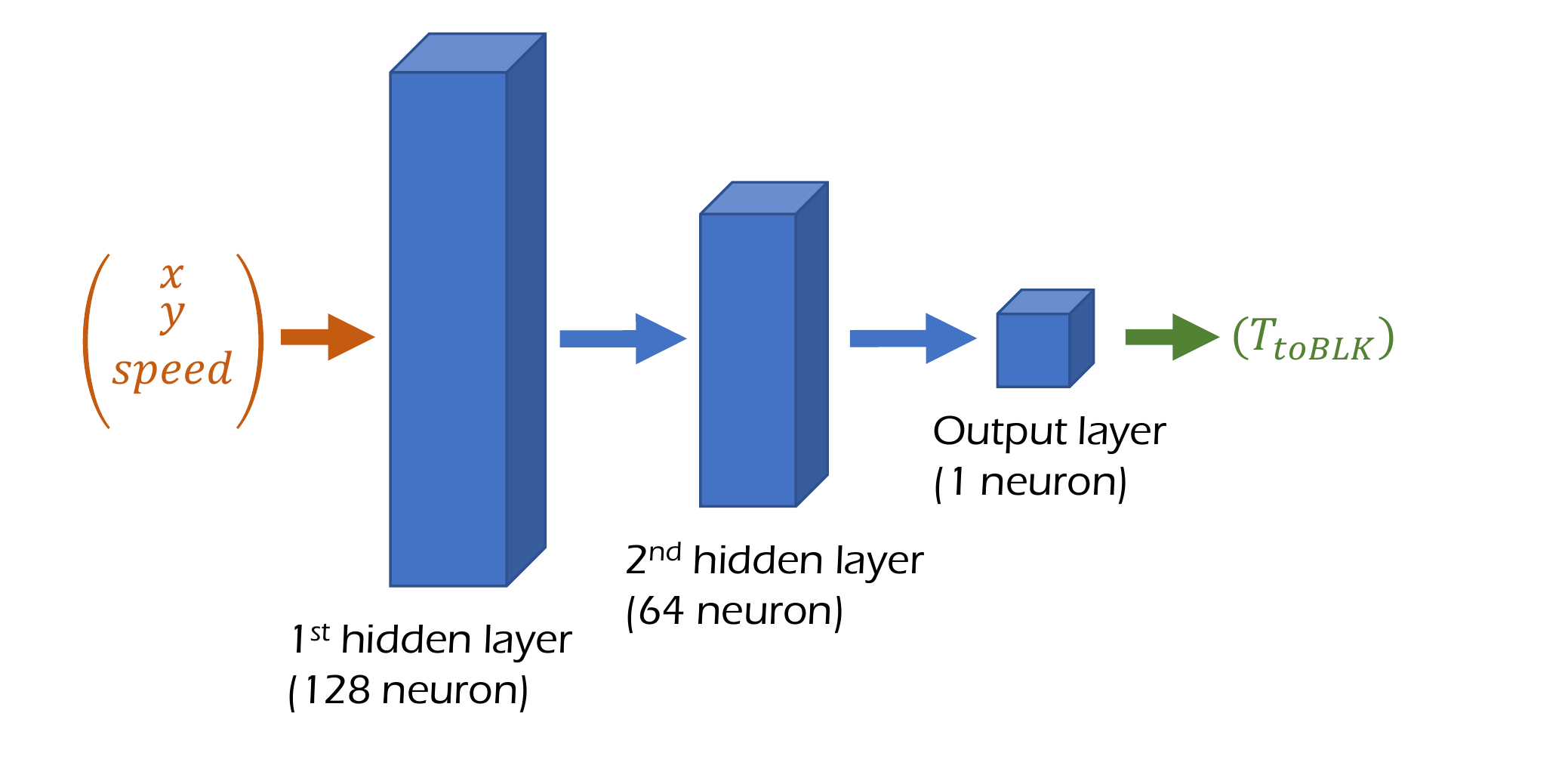}
\caption{A two-hidden layer neural network to perform regression.}
\label{NN}
\end{figure}

DL algorithms have achieved breakthroughs in various areas but at the expense of high computing and energy consumption. Combining CV with DL to assist the operation of UDNs will increase the models’ computational complexity, rendering this fusion inefficient \cite{[J.11]}. Since this study depends mainly on vision information, we carefully select the ML model that does not require much training time and achieves the expected results. In the previous section, we discussed using the pretrained YOLOv3 ODL model, which is off-the-shelf and can be used directly; thus, no further training is required. Moreover, to predict the $T_{toBLK}$, we select a simple NN model to conduct multivariate regression, as shown in Fig. \ref{NN}. The multivariate regression technique is a statistical approach that measures the relationship between dependent variables (i.e., $T_{toBLK}$) with more than one independent variable (i.e., x, y, speed). Besides, instead of using information-rich RGB images in model training/inferencing, the proposed technique only requires extracting the user’s location and speed to be fed into the NN model. This yields to significant savings in time.

\begin{table}
\centering
 \caption{Sample of the training dataset.}
 \includegraphics[width=0.4\textwidth]{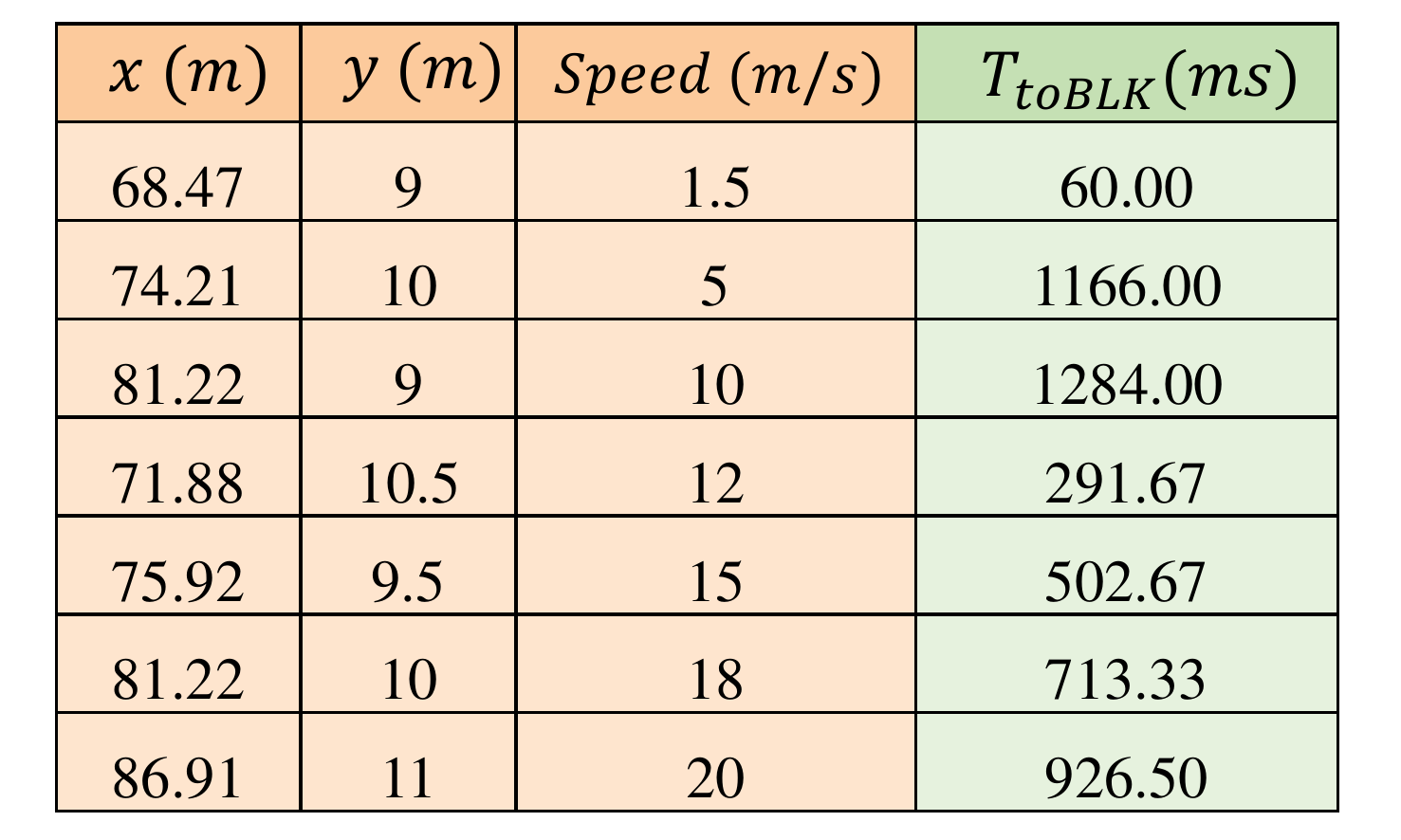}
 \label{sample}
\end{table}

\begin{figure}
\centering
\includegraphics[scale=0.37]{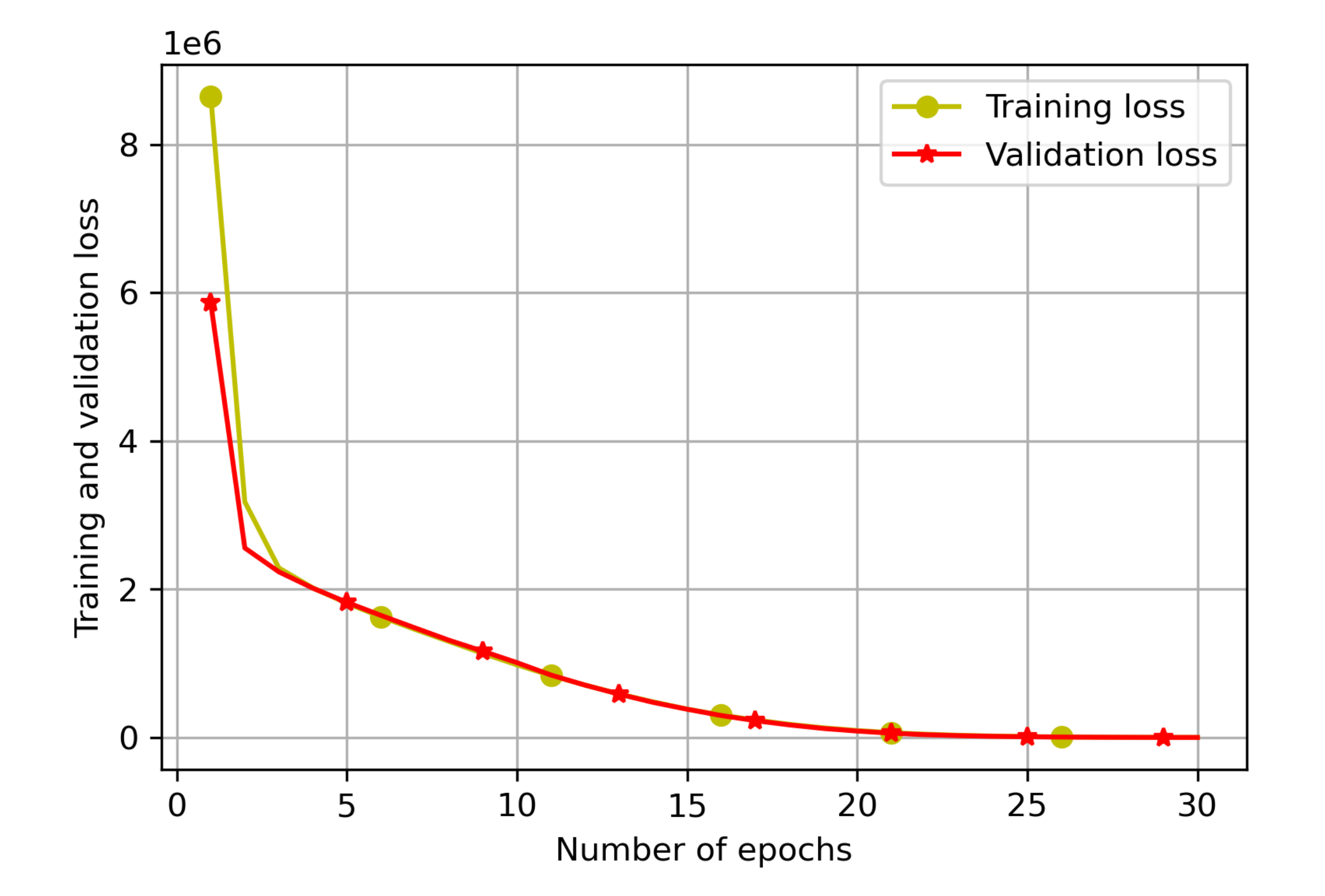}
\caption{Multivariate regression model training and validation loss versus number of epochs.}
\label{TVL}
\end{figure}

\paragraph{\textbf{Training Phase}}
The operations of the proposed regression model include model training and inference; both require the availability of data samples. For training the initial model, training datasets are readily generated using random (x, y) locations confined to our system model's street dimensions and using several speed values that reflect the expected vehicle speed in urban areas. The dependent variable $T_{toBLK}$ is calculated by fixing the location of the obstacle (bus) at an arbitrary location, for example at $x = 68.38m$ as illustrated in Fig. \ref{SM}, and using the speed formula given in (\ref{speed}). Table \ref{sample} shows a small sample of nearly ten thousand generated data samples that are divided into 80\% training, 20\% validation, and 10\% testing. Adaptive moment estimation (Adam) is used as an optimiser in the training process. Moreover, other model hyperparameters, such as the number of epochs, batch size, metric, and activation function, are set to E=50, B=20, mean square error (MSE), and linear activation function, respectively. Fig. \ref{TVL} shows the training and validation loss for the model in each epoch. This figure shows the effectiveness of this model since the training and validation loss approaches zero as the number of epochs increases. It is worth mentioning that the number of epochs depends on the number of data samples used for model training; a small dataset and a large number of epochs will lead to model overfitting, which is an unwanted behaviour for predictive modelling. The coefficient of determination (R squared) metric is adopted for model performance evaluation to measure the linear correlation between the predicted and actual values using the test dataset. The values of R squared range between 0 and 1; 1 is the optimal value that can be achieved. Using the test dataset, our model achieves 0.9998, which indicates the superb performance of this model. Finally, based on the generated dataset and the above mentioned hyperparameters, the proposed regression model took about 20 seconds to be trained using typical personal computer resources. However, using the MEC server, the training time will be less than one second \cite{[J.28]}.

\paragraph{\textbf{Inference Phase}}
Using the trained regression model, our framework is now ready to predict $T_{toBLK}$ for the user detected in images received from the cameras. After completing the street view and updating the blockage status, whenever the server receives RGB images from the SBSs, it will use the ODL to extract the location and speed of the moving user as discussed in Sec. \ref{ODETL}. The user’s information is now ready to be fed to the input layer of the regression model to infer the remaining time until the user reaches the blocked area. Furthermore, by using the same MEC server resources, the value of $T_{inf}$ is around 1 ms.

\subsection{Optimal Trigger Region Detection}\label{Opt_trigger}
Once all the parameters like user location, speed, and $T_{toBLK}$ are known, the final stage is to determine the optimal trigger region to perform a successful HO. This is the optimal distance at which the central server initiates the HO request after the BLK event is detected, and performs the PHO with minimum performance degradation. In this work, we used a threshold distance-based setup, where the central server determines the optimal trigger distance using the following equation:
\begin{equation}
    D = S_{u} \times T_{w},
    \label{TD}
\end{equation}
where $S_{u}$ is the speed of the user, which is already known from ODL. Using equation (\ref{TExec}), $T_{w}$ is given by 
\begin{equation}
    T_{w} = T_{exec} - T_{s},
    \label{T}
\end{equation}
where $T_{s}$ is the sum of other sub times i.e., $T_{RGB}$ few microseconds, $T_{ODL}$ 102 ms and $T_{inf}$ is 1 ms whereas the $T_{HO}$ is 50 ms as will be discussed in the next section. For successful HO, we assume $T_{toBLK}$ $\geq$ $T_{exec}$, and by substitution the above equation will become
\begin{equation}
    T_{w} = T_{toBLK} - T_{s}.
    \label{Td1}
\end{equation}

\begin{figure}
\centering
\includegraphics[scale=0.55]{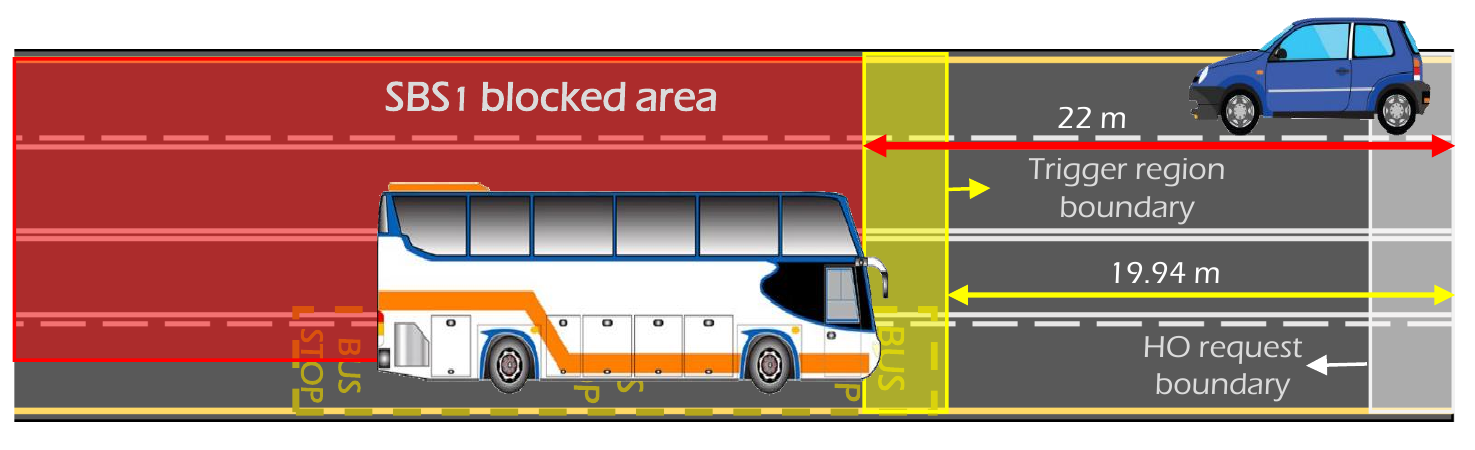}
\caption{Optimal trigger distance for a user with a speed of 30 mph.}
\label{OTR_SM}
\end{figure}

The optimal trigger region detection has two boundaries, i.e., the HO request boundary and the trigger region boundary, as shown in Fig. \ref{OTR_SM}. The HO request boundary is the point where the central server detects the BLK event and initiates the PHO request. On the other hand, the trigger region boundary, denoted by D, is the minimum distance from the blocked area where a successful HO is performed. For instance, if HO is performed well before the trigger region boundary, the user will experience an undesirable performance degradation due to the path loss. Therefore, the central server has to delay the HO for $T_{w}$ time, until the user reaches to optimal distance i.e., the boundary of the trigger region. To perform successful HO, the time $T_{s}$ is $153 ms$ which is fixed. The only variable parameter in optimal trigger boundary detection is $T_{w}$ that has a direct impact on the performance of the HO algorithm. For extensive analysis, optimal trigger distances based on different speeds are given in Table \ref{tab1}. For example, if a car is moving at the speed of 30 mph, the optimal trigger distance (D) to perform PHO is 19.94 m as shown in Fig. \ref{OTR_SM}. Once the BLK event is detected by the central server, the optimal trigger distance is calculated using equation (\ref{TD}), where $S_{u}$ is already known and $T_{toBKL}$ is obtained using regression analysis. It is possible to perform PHO at the HO request boundary. However, the early HO will have a significant degradation in RSS which is undesirable. Therefore, in our proposed model, the central server waits for $T_{w}$ time until the user reaches the optimal distance D to complete the PHO request.

\begin{table}[h]
\caption{Optimal trigger distance based on different speed of user.}
\begin{center}
\begin{adjustbox}{max width=\textwidth}
\begin{tabular}{|c|c|c|c|}
\hline
\text{Speed (mph)} & \text{$T_{toBLK}$ (sec)} & \text{$T_{w}$ (sec)} & \text{D (m)}\\
\hline
5  & 9.85 & 9.66 & 21.67\\
10 & 4.92 & 4.76 & 21.30\\
15 & 3.28 & 3.127 & 20.96\\
20 & 2.46 & 2.30 & 20.62\\
25 & 1.97 & 1.82 & 20.19\\
30 & 1.64 & 1.487 & 19.94\\
35 & 1.40 & 1.25 & 19.58\\
\hline
\end{tabular}
\end{adjustbox}
\label{tab1}
\end{center}
\end{table}

\begin{figure*}
\centering
\includegraphics[width=1.02\textwidth]{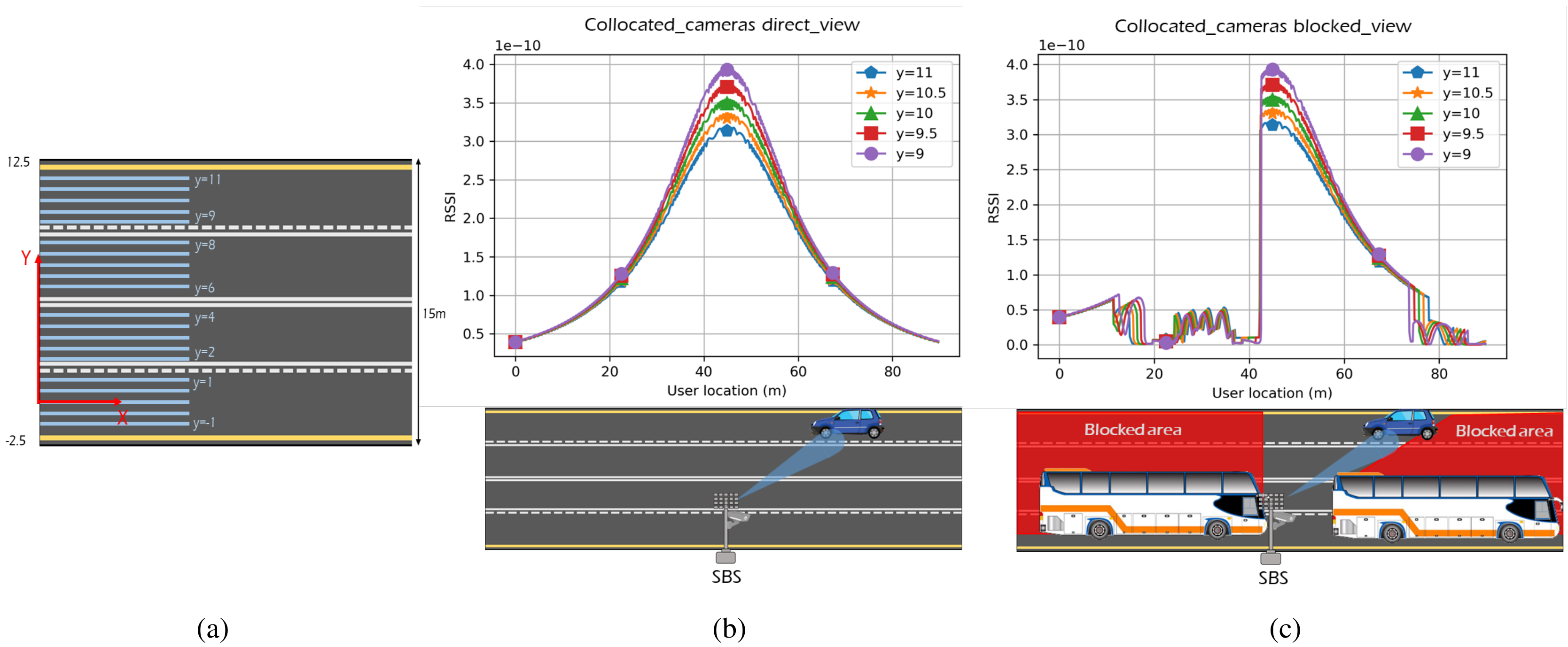}
\caption{Analysing ViWi information to construct the system model. (a) Locating the origin of the Cartesian coordinates, (b) Using ViWi information from colocated cameras direct view scenario to model SBS$_{2}$ in our system model, and (c) ViWi information from colocated cameras blocked view scenario shows a similar RSS pattern when there is no blockage.}
\label{Analysis}
\end{figure*}

\subsection{Proactive Handover Mechanism}
$T_{HO}$ is one of the key parameters of our algorithm to determine whether there is enough time to perform HO and avoid radio links failures. It is shown that the time needed to complete the HO process in long term evolution (LTE) networks is 50 ms \cite{[J.12]}. Since the HO procedure in new radio (NR) is similar to that in LTE networks \cite{[J.12]}, we assume that $T_{HO}$ equals 50 ms. Since the values of all parameters in (\ref{TExec}) are determined, the central server is aware of the time needed to execute the proposed algorithm, $T_{exec}$ equals (153 + $T_{w}$) ms. If the central server detects a BLK event in the received RGB images, the server will predict the time needed until the user reaches the blocked area ($T_{toBLK}$). If $T_{toBLK}$  is greater than $T_{exec}$, then our algorithm has a high probability of successfully triggering and completing HO. Whereas when $T_{toBLK}$ is less than $T_{exec}$, the time needed to complete the HO process and avoid radio link failure is insufficient, which means that the user undergoes a service interruption.

\section{PERFORMANCE EVALUATION AND RESULTS}\label{PEval}
To investigate the effectiveness of the proposed CV-based PHO framework, we utilise a publicly available dataset called vision wireless (ViWi) \cite{[J.16]}. The ViWi dataset combines visual and wireless information generated using Wireless InSite ray-tracing software and 3D game modelling for mmWave wireless systems. It considers four different scenarios based on the camera location (collocated and distributed) and the view (direct and blocked). Furthermore, each data sample represents 4-tuple of user location, RGB image, depth image, and the wireless channel. The distinctive feature of ViWi is that it is a parametric, systematic, and scalable data generation framework that can be used to produce data based on different scenario requirements. In the performance evaluation, the focus is to track the strength of the signal received by the moving user and test the value of the proposed algorithm in maintaining a good received signal during movement compared to traditional mmWave systems (i.e., without PHO).

\subsection{Simulation Setup}
\begin{figure}
\centering
\includegraphics[scale=0.38]{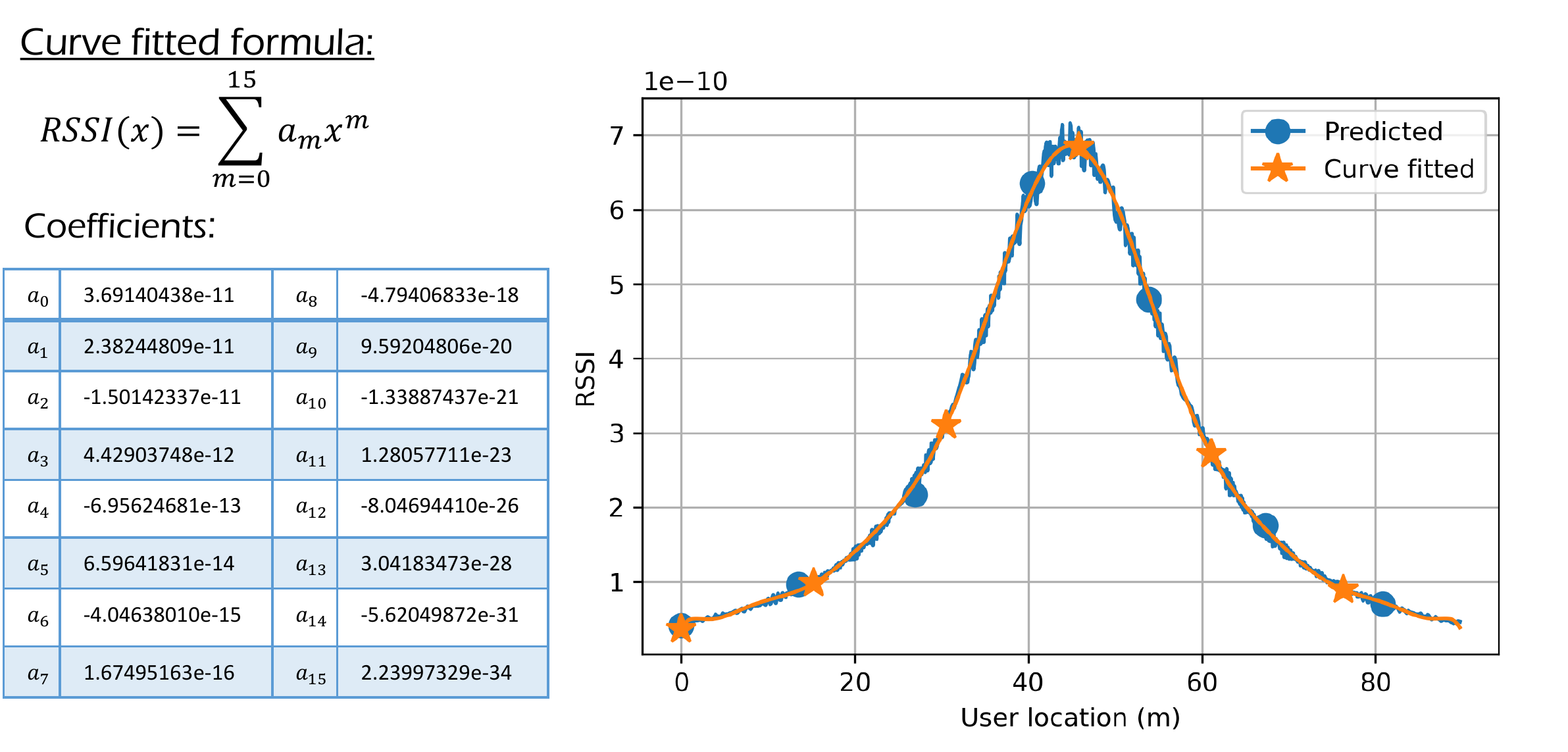}
\caption{Determining the RSS from SBS$_{2}$ at trajectory \textit{y}= 9 using the curve fitting tool.}
\label{PF}
\end{figure}

\vspace{-0.05cm}
In our simulation, we consider a simple environment containing a blocking object located near the SBS$_{1}$ and a single user moving at a speed of 30 mph, as illustrated in Fig. \ref{SM}. Our system model is different from any of the scenarios introduced with the ViWi dataset; however, we were able to produce both visual and wireless data for our model by merging the two scenarios of collocated cameras, direct and blocked view. Fig. \ref{Analysis} shows the analysis performed to form our system model. Initially, we analyse the information provided in the ViWi dataset to determine the place of origin of the Cartesian coordinate system, as it is not mentioned in the ViWi dataset. We used the ViWi trajectories (\textit{y}= 9 to 11) to identify the location of the origin and other trajectories, as shown in Fig. \ref{Analysis}(a). Then, for each trajectory considered in ViWi, we plot the RSS against the vehicle location for the two scenarios, as illustrated in Fig. \ref{Analysis} (b) and (c). From these figures, we can conclude that each SBS has the same signal pattern (i.e. bell shape) in any trajectory with the highest signal strength at the same x-location of the SBS. To model the RSS of SBS$_{2}$, we used the information in Fig. \ref{Analysis}(b) and applied curve fitting to generate the mathematical formula. First, we predict the RSS value for each other trajectory (\textit{y}= -1 to \textit{y}= 8) at each user location, and then we used the curve fitting process to create the mathematical formula that gives the RSS versus vehicle location (x). Fig. \ref{PF} illustrates the generated mathematical formula and RSS from SBS$_{2}$ at \textit{y}=9, which is the trajectory used in the evaluation. Finally, Python programs are used to conduct the simulations.

\subsection{Simulation Results}
\begin{figure}
\centering
\includegraphics[scale=0.55]{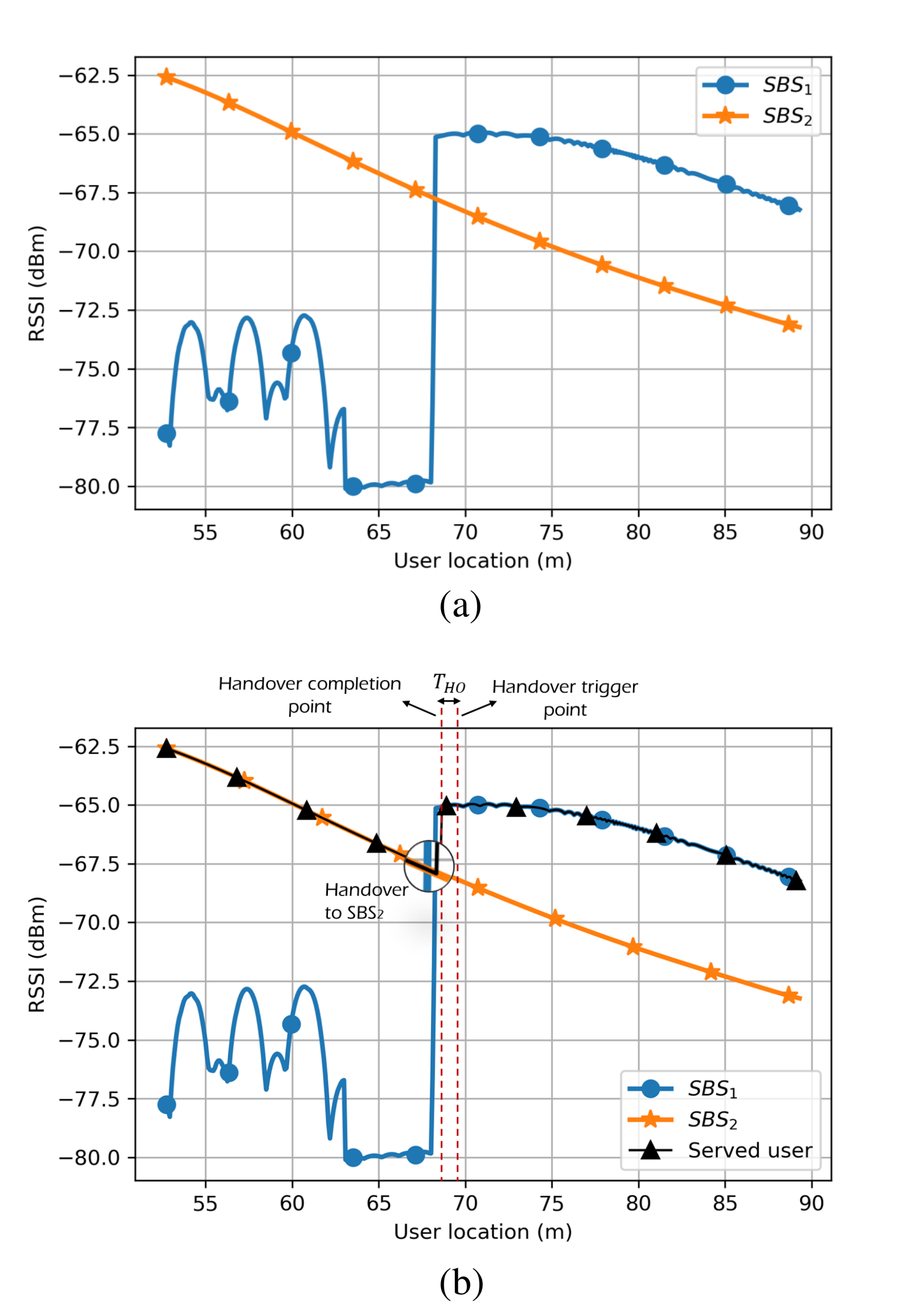}
\caption{Performance evaluation of the proposed framework. (a) RSS from SBS$_1$ and SBS$_2$, (b) Using the CV-assisted PHO algorithm to detect BLK event and trigger PHO.}
\label{combined}
\end{figure}

In what follows, we examine the usefulness of the proposed algorithm in maintaining physical link connectivity and ensuring a timely and seamless transition from one SBS to another. The RSS indicator (RSSI) is used as a metric to measure the quality of the received signals from the nearby SBSs. Fig. \ref{combined}(a) plots the RSSI received from SBS$_1$ and SBS$_2$ at the \textit{y}=9 trajectory along the street. It is noticed that the signal drops from SBS$_1$ when the user reaches the area behind the blocking object since the beam undergoes severe attenuation. In contrast, the signal received from  SBS$_2$ does not experience any interruptions because the LoS path is clear between the user and the SBS$_2$. In a traditional wireless network (i.e., does not employ the PHO algorithm), the vehicle experiences a connectivity disruption when entering the blocking area. Such interruption may lead to service drop and the need to initiate a new connection, which means more delay and poor QoE, inconsistent with the vision of 5G/6G wireless networks of providing ultra reliable and low latency communications. Fig. \ref{combined}(b) demonstrates the capability of the proposed PHO algorithm in proactively predicting beam blockages. This figure reveals the efficiency of the PHO algorithm in identifying the BLK event in advance and triggering a timely HO. The merit of determining the optimal point of triggering HO in our algorithm is vital to maintain the QoE at high levels and avoid early HO, which could lead to bad system performance. The points of triggering HO and HO completion are also shown in the figure, which illustrates how our algorithm is also QoE-aware. 

\begin{figure}
\centering
\includegraphics[width=\linewidth]{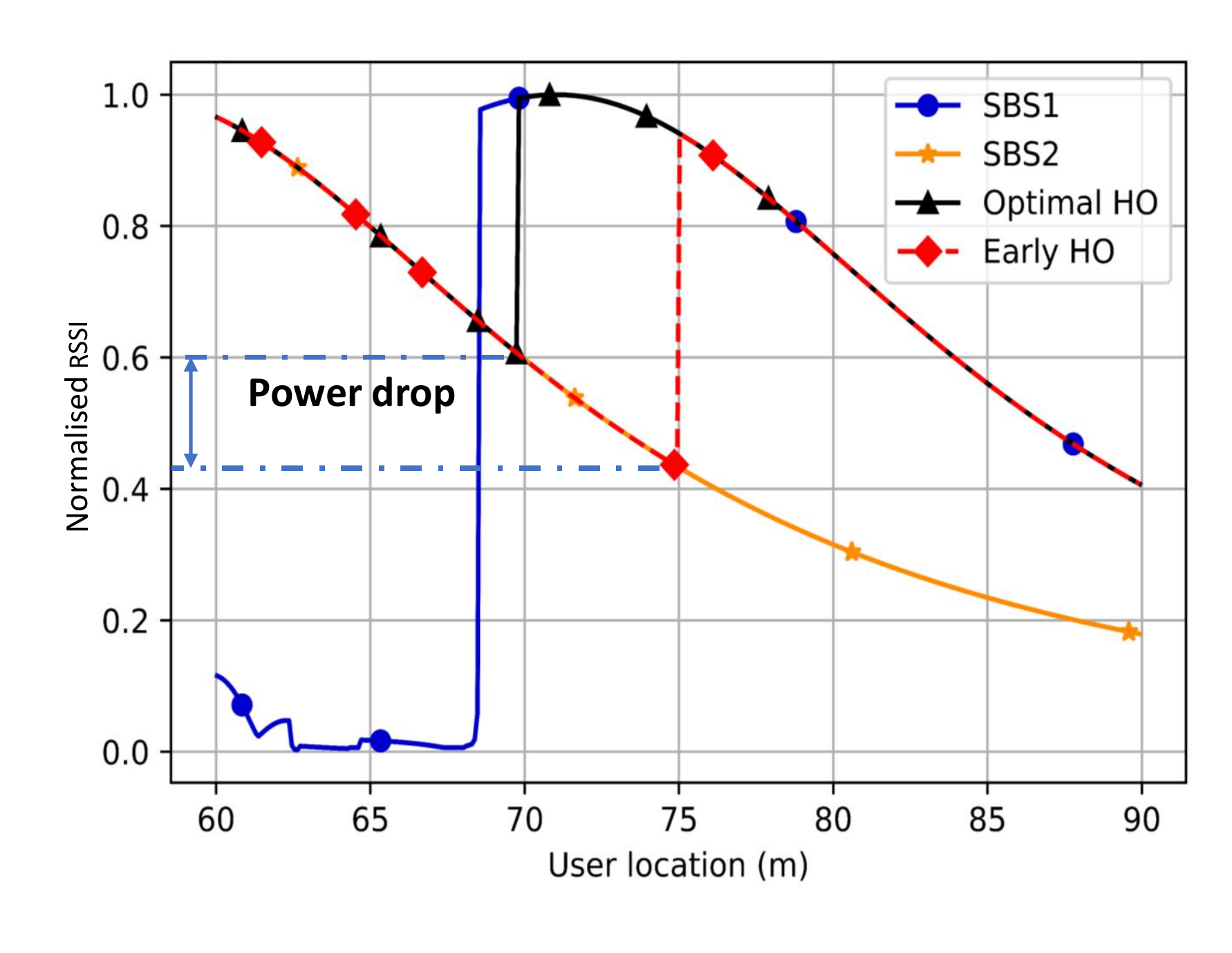}
\caption{The normalised RSSI as function of the user location when the user speed is fixed at 30 mph.}
\label{OTR}
\end{figure}

Fig. \ref{OTR} shows the results of the optimal trigger region for the user moving at the speed of 30 mph. The boundary of the optimal trigger region is the minimum distance where the central server can perform the successful HO with minimum performance degradation measured in the percentage drop in normalised RSSI. In our case, the optimal distance for the user moving at a speed of 30 mph is found to be around 70 m from origin. The central server performs HO once a BLK event is detected, as a result, the resources of the user shift from SBS1 to SBS2. During the HO process, the user will experience a drop in RSSI due to path loss. For instance, if the central server performs an early HO i.e., 5 m before the trigger region boundary, there is a power drop of approximately 20 \% as shown in Fig. \ref{OTR}. therefore, the optimal trigger distance provides the trade-off between the PHO success rate and the drop in RSSI to maintain the seamless connectivity. 

\begin{figure}
\centering
\includegraphics[scale=0.4]{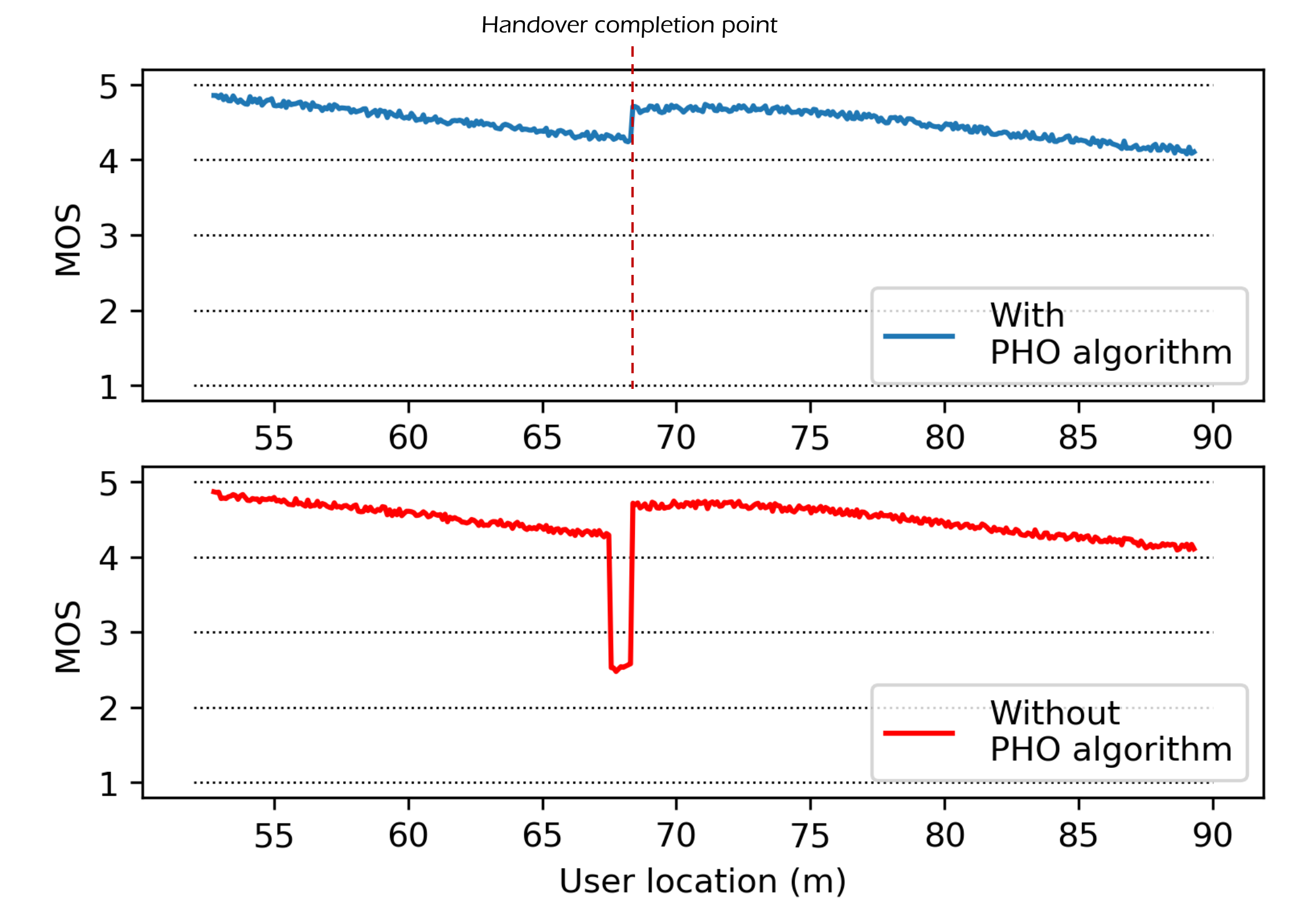}
\caption{Measuring the QoE with and without PHO.}
\label{MOS}
\end{figure}

Next, we investigate the effectiveness of the PHO algorithm in improving the reliability of UDN by considering a real-time application sensitive to service interruption and network latency. We assume a moving user that is running a video call and consider the mean opinion score (MOS) as a metric, which is a measurement of the QoE. MOS is a measure of the media's overall perceived service quality based on human judgement and ranges from 1 to 5 (1-bad, 2-poor, 3-fair, 4-good, 5-excellent) \cite{[J.29]}. Fig. \ref{MOS} shows the MOS value versus the user location, which is experiencing different RSS based on the distance from the SBS and the presence of obstacles. To translate the values of RSS to the corresponding values of MOS, we adopt the mapping table in \cite{[J.17]}. From Fig. \ref{MOS} we notice that without the PHO algorithm, the user will experience a service interruption when it reaches the blocked area, and the call quality is dropped by 40\% to the poor MOS region. Call quality will remain poor until the user drops the serving SBS link and connects to a new SBS. Whereas, using the PHO technique, our algorithm intelligently detects the existence of a blockage and countermeasures the possible signal blockage by triggering HO in advance. Therefore, maintaining the perceived MOS at the excellent region. This work enhances the reliability of UDNs that will facilitate the realisation of future latency-sensitive applications.

\section{CONCLUSIONS} \label{Conclusion}
This paper proposed a novel CV-based PHO framework to address the challenge of frequent HO and beam blockage in UDNs. The key idea is to raise the network's awareness of the surrounding environment by leveraging visual information and using the CV to predict BLK events and perform PHO. A pretrained object detection model, in addition to a multivariate regression model, are used to predict the obstacle/user location and the time remaining before the user arrives at the blocked area. Moreover, this framework is QoE-aware, where we presented an analysis of the optimal location/time to perform HO while minimising the drop in QoE. The evaluation results demonstrated that this framework is able to avoid 40\% service reduction and maintain a high level of perceived QoE. Accordingly, this work improves the performance of the UDN by making it more dynamic in interaction with the environment, which is inline with the vision of achieving low-latency and time-sensitive applications in B5G and 6G networks.

\balance
\bibliographystyle{IEEEtran}
\bibliography{PHO}

\end{document}